\UseRawInputEncoding
\documentclass[11pt,aps,amssymb,prd,a4paper,nofootinbib]{revtex4-1}
\usepackage{fullpage}
\usepackage{amsfonts}
\usepackage{amsmath}
\usepackage{slashed}
\usepackage{amssymb}
\usepackage{graphicx}
\usepackage{makeidx}
\usepackage{cancel}
\usepackage{epic}
\usepackage{eepic}
\usepackage{epsfig}
\usepackage{latexsym}
\usepackage[dvipsnames]{xcolor}
\usepackage{float}
\usepackage{multirow}
\usepackage[export]{adjustbox}
\usepackage{xurl,hyperref}
\usepackage{enumitem}
\hypersetup{colorlinks=true,citecolor=red,linkcolor=NavyBlue,urlcolor=NavyBlue}
\usepackage[utf8]{inputenc}
\usepackage[caption=false]{subfig}
 
\usepackage{natbib}
\usepackage{relsize}
\usepackage[left=2.2 cm,right=2.2 cm,top=2 cm,bottom=2 cm]{geometry}
\usepackage{mathptmx}

\begin{document}
\relscale{1.05}
\captionsetup[subfigure]{labelformat=empty}

\title{The Revival of $U(1)_{L_e-L_\mu}$: A Natural Solution for $(g-2)_\mu$ with a Sub-GeV Dark Matter}

\author{Bibhabasu De}
\email{bibhabasude@gmail.com}
\affiliation{Department of Physics, The ICFAI University Tripura, Kamalghat-799210, India}

\date{\today}

\begin{abstract}
\noindent
The experiments searching for a hidden gauge sector with a leptophilic neutral gauge boson have already ruled out $U(1)_{L_e-L_\mu}$ as a feasible extension of the Standard Model~(SM) gauge group~($\mathcal{G}_{\rm SM}$) for explaining the observed discrepancy in $(g-2)_\mu$. The paper proposes a simple extension of the minimal particle content of $\mathcal{G}_{\rm SM}\otimes U(1)_{L_e-L_\mu}$ with a TeV-scale scalar leptoquark $S_1$. Due to the non-trivial transformation of $S_1$ under $U(1)_{L_e-L_\mu}$, the model generates additional one-loop contributions to $(g-2)_\mu$, reviving the considered gauge extension within the experimentally allowed regions of the parameter space. The model can also accommodate a viable Dark Matter~(DM) candidate $\chi$ --- a vector-like SM-singlet fermion in the sub-GeV mass regime. The theory provides a natural framework to test the proposed DM phenomenology through electron excitation signals. Moreover, the DM-specific observables and $(g-2)_\mu$ being connected through the {\it New Physics}~(NP) parameters, the future beam dump experiments hunting for light, feebly interacting particles and the DM direct detection experiments are complementary to each other for constraining/falsifying the model.    
 \end{abstract}
	
\maketitle	

\section{Introduction} 
\noindent
The Standard Model has been tested to an exceptional degree of precision at the high-energy colliders and through various other experiments, establishing itself as the {\it nature's} theory for strong and electroweak~(EW) interactions. The recent examples include the discovery of 125 GeV Higgs boson at the Large Hadron Collider~(LHC)~\cite{ATLAS:2012yve,CMS:2012qbp}, and the remeasurement of the W-boson mass at the ATLAS~\cite{ATLAS:2024erm} and CMS~\cite{CMS-PAS-SMP-23-002}, where SM predictions closely match with the experimental data. However, despite all its excellence as a theoretical framework, SM falls short to explain certain observations, e.g., the astrophysical and cosmological signatures of dark matter~\cite{1932BAN.....6..249O,Zwicky:1933gu, Zwicky:1937zza,Metcalf:2003sz,Planck:2018vyg}, the discrepancy between the observed and predicted values of the anomalous magnetic moments of muon~\cite{Muong-2:2023cdq} and electron~\cite{Parker_2018,Morel:2020dww}, neutrino oscillations~\cite{Pontecorvo:1967fh}, etc. In the SM, particularly, the anomalous magnetic moment of muon~$\left({\rm i.e.,}~(g-2)_\mu/2\right)$ has been predicted with a  significant precision, including the EW and hadronic contributions~\cite{Aoyama:2020ynm}, while experiments have also measured it very precisely. Thus, $(g-2)_\mu$ is a crucial parameter to test the precision of the SM at the quantum level. However, the observations indicate a non-negligible difference between the experimentally measured and predicted values of $(g-2)_\mu$~\cite{Muong-2:2023cdq}. Though a similar discrepancy has been observed for the electrons, measurements using the recoil on Cs~\cite{Parker_2018} and Rb~\cite{Morel:2020dww} atoms result in a relative sign between the reported values. Definitely, such observations strengthen the possibility of having a theory beyond the standard model~(BSM). However, the NP can only exist either at a higher energy scale or has to be weakly/selectively coupled to the SM fields to be consistent with the current experimental results. From the theoretical perspective, a quite natural and well-motivated way to construct a BSM framework is to extend the SM gauge group with an additional $U(1)$ symmetry~\cite{Langacker:2008yv}. After spontaneous symmetry breaking~(SSB), the NP interactions may arise through the associated neutral gauge boson $Z^\prime$, creating a room for various BSM observables that can't be accommodated within the parameter space of the SM. Such abelian extensions of the SM can originate from Grand Unified Theories~(GUT)~\cite{Robinett:1982tq,Langacker:1984dc}, extra-dimensional models~\cite{Antoniadis:1990ew,Appelquist:2000nn} or string compactifications~\cite{Goodsell:2010ie}. However, $U(1)-$augmentations of $\mathcal{G}_{\rm SM}$ can also be formulated from the accidental global symmetries of the SM. Classically, the SM Lagrangian features the global symmetry group $U(1)_B\otimes U(1)_{L_e}\otimes U(1)_{L_\mu}\otimes U(1)_{L_\tau}$, ensuring the conservation of baryon number $B$ and the individual lepton numbers $L_i$~[$i=e,\,\mu,\,\tau$]. The difference between any two lepton numbers, i.e., $L_i-L_j$~[$i,\,j=e,\,\mu,\,\tau$ with $i\neq j$] can naturally be promoted to a gauge quantum number as the associated abelian group $U(1)_{L_i-L_j}$ represents an anomaly-free theory even in its minimal form~\cite{Foot:1990mn,He:1991qd,Foot:1994vd}. These theories, with $Z^\prime$ as the only NP field, provide a simple and direct explanation for the observed discrepancies in lepton $(g-2)$ --- $U(1)_{L_\mu-L_\tau}$ and $U(1)_{L_e-L_\tau}$ being specific to $(g-2)_\mu$ and $(g-2)_e$, respectively, whereas $U(1)_{L_e-L_\mu}$ is a {\it good} theory for both of the discrepancies. Note that, the neutral gauge bosons of $U(1)_{L_i-L_j}$ groups are automatically leptophilic with only loop-induced couplings to the quark sector, making them tough to constrain through the hadronic colliders\footnote{The phenomenological implications of a flavored $Z^\prime$ have been studied in Ref.~\cite{Chun:2018ibr}.}. However, the parameter spaces can be deeply probed through other experiments~\cite{Riordan:1987aw,Bjorken:1988as, Davier:1989wz, Bross:1989mp, Bjorken:2009mm, Andreas:2012mt,APEX:2011dww,A1:2011yso, NA64:2016oww,CHARM:1985nku,LSND:1997vqj,Blumlein:2011mv, Blumlein:2013cua,BaBar:2009lbr, BaBar:2014zli,KLOE-2:2011hhj,KLOE-2:2012lii,KLOE-2:2016ydq,Anastasi:2015qla,Kaneta:2016uyt,Bilmis:2015lja,Lindner:2018kjo,Super-Kamiokande:2011dam,Ohlsson:2012kf,Gonzalez-Garcia:2013usa,Dreiner:2013tja} leading to an extremely constrained scenario in the $M_{Z^\prime}-g^\prime$ plane~(a complete analysis can be found in Ref.~\cite{Bauer:2018onh}). Here, $M_{Z^\prime}$ denotes the mass of $Z^\prime$ in the broken phase of $U(1)_{L_i-L_j}$, and $g^\prime$ stands for the corresponding gauge coupling. To be specific, only a tiny region in the parameter space of $U(1)_{L_\mu-L_\tau}$ is still available for explaining the observed discrepancy in muon anomalous magnetic moment~($\Delta a_\mu$), whereas $U(1)_{L_e-L_\mu}$ has been completely ruled out as a viable explanation for $\Delta a_\mu$ and $\Delta a_e$~\cite{Bauer:2018onh,Greljo:2022dwn}. A similar conclusion goes for $U(1)_{L_e-L_\tau}$. The present paper proposes an economical extension of the $\mathcal{G}_{\rm SM}\otimes U(1)_{L_e-L_\mu}$-particle spectrum with a TeV-scale scalar leptoquark~(LQ) such that $U(1)_{L_e-L_\mu}$ can be revived as a pursuable BSM theory to explain the observed $\Delta a_\mu$. However, note that the framework is equally adaptable for all the three $U(1)_{L_i-L_j}$ extensions of the SM gauge group.

Leptoquarks~(for a recent review, see Ref.~\cite{Dorsner:2016wpm}) are hypothetical bosons charged under $U(1)_B\otimes U(1)_{L_e}\otimes U(1)_{L_\mu}\otimes U(1)_{L_\tau}$ in the classical sense. However, their origin is rooted within the GUTs~\cite{Pati:1974yy, Georgi:1974sy, Georgi:1974my} as a mediator of the quark-lepton interaction, making them a vital candidate for a plethora of BSM theories. For example, several $B$-anomalies can be explained by extending the SM with a LQ~\cite{Dorsner:2013tla,Gripaios:2014tna,Becirevic:2015asa,Becirevic:2016yqi, Crivellin:2017zlb,Cline:2017aed,DiLuzio:2017chi,Mandal:2018kau,Aydemir:2019ynb,Crivellin:2019dwb,Asadi:2023ucx}. LQs can also play an important role in explaining the charged lepton flavor violating~(CLFV) processes~\cite{PhysRevD.93.015010,Mandal:2019gff,Crivellin:2020mjs,De:2024foq}, DM phenomenology~\cite{Mandal:2018czf,Choi:2018stw,Mohamadnejad:2019wqb} and the production of scalars at the high-energy colliders~\cite{Bhaskar:2020kdr,Bhaskar:2022ygp,DaRold:2021pgn,Agrawal:1999bk,Enkhbat:2013oba,De:2024tbo}. Moreover, a minimal extension of the SM with a LQ can easily produce a significant BSM contribution to $(g-2)_\mu$~\cite{Djouadi1990,Cheung:2001ip,ColuccioLeskow:2016dox,Dorsner:2019itg,Mandal:2019gff,De:2023acg}. This is crucial for the proposed objective as the $\mathcal{G}_{\rm SM}\otimes U(1)_{L_e-L_\mu}$ theory contributes subdominantly to $(g-2)_\mu$ in the allowed parameter space. Note that, in the simplest GUT models, LQs acquire mass at a scale far beyond the reach of the current and future colliders. However, there exist GUTs where the stability of protons can be explained with a TeV-scale scalar LQ~\cite{BUCHMULLER1986377,Murayama:1991ah,Dorsner:2005fq, GEORGI1979297, FileviezPerez:2007bcw, Senjanovic:1982ex,Dorsner:2004jj, Aydemir:2019ynb, Dorsner:2024seb}. Therefore, in principle, the gauge group $\mathcal{G}_{\rm SM}\otimes U(1)_{L_e-L_\mu}$ has to follow from two different symmetry breaking chains: $\mathcal{G}_{\rm GUT}\to \mathcal{G}_{\rm SM}$ and $SU(2)_{L_e-L_\mu}\to U(1)_{L_e-L_\mu}$~\cite{He:1991qd, Bauer:2018onh}, where $\mathcal{G}_{\rm GUT}$ can be any possible grand unifying formulation allowing for the TeV-scale interactions of a scalar LQ.

An added advantage of the considered gauge extension is the DM phenomenology. The existence of DM has already been confirmed through its gravitational interactions~(for review, see Refs.~\cite{Bertone:2004pz,Jungman:1995df,GONDOLO1991145,Cirelli:2024ssz}), while the Cosmic Microwave Background~(CMB) anisotropy provides a concrete estimation for its abundance~\cite{Planck:2018vyg}. However, its particle nature is still unknown and can't be explained with any of the SM fields. The $U(1)$-extensions of the SM present a simple framework where the assumed DM candidate interacts with the SM particles through $Z^\prime$ to produce a correct relic density. Following the same line, a vector-like SM singlet fermion can be introduced in the proposed model as a viable DM candidate, which, being non-trivially charged under $U(1)_{L_e-L_\mu}$ interacts with the SM leptons through the $Z^\prime$ portal. Vector-like fermions are theoretically well-motivated BSM candidates~\cite{Hewett:1988xc, Langacker:1980js, Antoniadis:1990ew, Minkowski:1977sc, Foot:1988aq} to construct anomaly-free UV-complete theories. Due to the leptophilic nature of $Z^\prime$, the considered framework is particularly suitable for describing a sub-GeV DM as the mass regime is kinematically more sensitive to DM-electron scattering than the nuclear recoils. However, electrons being a bound state, the kinematics differs significantly from that of the nuclear scattering events. The direct relation between momentum transfer and recoil energy is no longer valid for the electronic recoils. If $\Delta E_{\rm DM}$ is the energy lost by the DM due to scattering on an atomic target, a part of it is used for overcoming the binding energy~($E_B^k$) of the $k^{\rm th}$ electron. Thus, only $\Delta E_{\rm DM}-E_B^k$ appears as the recoil energy of the electron~(since atomic recoil is negligible). Further, for semiconductor targets, the electron wavefunctions are highly delocalized over the entire lattice, and one has to adopt sophisticated numerical techniques to compute the DM scattering rate. However, due to their low thresholds, semiconductor detectors are very useful for probing the sub-GeV mass regime. The present paper considers Si and Ge as the target materials for studying the detection prospects of the proposed DM candidate through electron excitation signals. For earlier works with a similar approach to the DM direct detection, refer to Refs.~\cite{Essig:2011nj,Graham:2012su,Essig:2015cda,Foldenauer:2018zrz,Dutta:2019fxn,Knapen:2021run,Hochberg:2021pkt,Kahn:2021ttr,Chen:2022xzi,Barman:2024lxy}. 

The rest of the paper is organized as follows. Sec.~\ref{sec:model} establishes the BSM formulation based on $\mathcal{G}_{\rm SM}\otimes U(1)_{L_e-L_\mu}$ gauge theory with each NP field chosen for a specific purpose. The DM phenomenology has been studied in Sec.~\ref{sec:DM}. Sec.~\ref{sec:mu} presents the most significant part of this work --- it has been explicitly shown that the entire allowed parameter space of $\mathcal{G}_{\rm SM}\otimes U(1)_{L_e-L_\mu}$ can be availed for explaining $\Delta a_\mu$ in the presence of a TeV-scale scalar LQ. Finally, the paper has been concluded in Sec.~\ref{sec:conc}.

\section{The Model}
\label{sec:model}
\noindent
The model considers a simple extension of the SM gauge group $\mathcal{G}_{\rm SM}=SU(3)_C\otimes SU(2)_L\otimes U(1)_Y$ with $U(1)_{L_e-L_\mu}$. As discussed in the Introduction, the difference between electron and muon lepton numbers, i.e., $L_e-L_\mu$, can be naturally gauged to an anomaly-free theory without assuming any extra fermion field. However, to explain the observed discrepancy in $(g-2)_\mu$, one has to augment the minimal particle spectrum of the assumed gauge theory. A phenomenologically well-motivated candidate can be the scalar leptoquark $S_1$ as it directly couples the charged leptons to the up-type quarks and can be embedded into a UV complete theory supporting its TeV-scale interactions~\cite{Dorsner:2004jj, Aydemir:2019ynb, Dorsner:2024seb}. The non-zero $U(1)_{L_e-L_\mu}$ charge of $S_1$ will ensure that it couples only to the $2^{\rm nd}$ generation leptons resulting in an additional BSM contribution to $(g-2)_\mu$ at the one-loop level. Further, as a consequence of $S_1$ being charged under $U(1)_{L_e-L_\mu}$, the non-observation of CLFV processes can be trivially explained. Moreover, the DM phenomenology can also be described within the theory by introducing a vector-like SM-singlet fermion $\chi$ without disturbing the anomaly-free structure of the minimal model. The spontaneous breaking of $U(1)_{L_e-L_\mu}$ symmetry can be achieved through an SM-singlet scalar $\phi$ which carries a non-trivial $U(1)_{L_e-L_\mu}$ charge $Q_\phi$. Note that, for an unbroken $U(1)_{L_e-L_\mu}$, the absolute stability of $\chi$ could be warranted with its non-zero abelian charge $Q_\chi$. But for the considered framework one has to impose an extra $Z_2$ symmetry to stabilize it as a proposed DM candidate. All other fields are considered even under this additional discrete symmetry. Thus, the extended $\mathcal{G}_{\rm SM}\otimes U(1)_{L_e-L_\mu}$ model augments the SM particle content with three BSM fields --- $S_1$, $\chi$, $\phi$. Table~\ref{tab:parti} shows the full list of particles and their transformations under $\mathcal{G}_{\rm SM}\otimes U(1)_{L_e-L_\mu}$. 
\begin{table}[!ht]
\begin{tabular}{|c|c|}
\hline
Fields & $SU(3)_C\otimes SU(2)_L\otimes U(1)_Y\otimes U(1)_{L_e-L_\mu}$ \\
\hline
\hline
$L_L^1=(\nu_e\quad e)^T$  & ({\bf 1}, {\bf 2}, $-1/2$, 1)  \\
$L_L^2=(\nu_\mu\quad \mu)^T$  & ({\bf 1}, {\bf 2}, $-1/2$, $-1$)  \\
$L_L^3=(\nu_\tau\quad \tau)^T$ & ({\bf 1}, {\bf 2}, $-1/2$, $0$)  \\
		$ e_R$ & ({\bf 1}, {\bf 1}, $-1$, 1) \\
		$\mu_R$ & ({\bf 1}, {\bf 1}, $-1$, $-1$) \\
		$\tau_R$ & ({\bf 1}, {\bf 1}, $-1$, $0$) \\
		$Q_L=(u_L\quad d_L)^T$ & ({\bf 3}, {\bf 2}, $1/6$, 0) \\
		$U_R=(u_R,\,c_R,\,t_R)$ & ({\bf 3}, {\bf 1}, $2/3$, 0) \\
		$D_R=(d_R,\,s_R,\,b_R)$ & ({\bf 3}, {\bf 1}, $-1/3$, 0) \\
		$H = (H^+ \quad H^0)^T $ & ({\bf 1}, {\bf 2}, $1/2$, 0)  \\
\hline
		$S_1$ & ($\mathbf{\bar{3}}$, {\bf 1}, $1/3$, $1$)	\\
		$\phi$ & ({\bf 1}, {\bf 1}, 0, $Q_\phi$)	\\
		$\chi_{L,\,R}$ & ({\bf 1}, {\bf 1}, 0, $Q_\chi$)	\\
		\hline
\end{tabular} 
\caption{Fields and their transformations under $\mathcal{G}_{\rm SM}\otimes U(1)_{L_e-L_\mu}$. The electromagnetic~(EM) charge of a particle has been defined as $Q_{\rm EM}=T_3+Y$.}
\label{tab:parti}
\end{table}

The complete Lagrangian for the proposed framework can be cast as,
\begin{align}
\mathcal{L}=\mathcal{L}_{\rm SM}+\mathcal{L}_{\rm NP}-\mathbb{V}(H,\,S_1,\,\phi)~,
\end{align}
where $\mathcal{L}_{\rm SM}$ corresponds to the interactions of the SM fields only, while all the NP interactions are encapsulated in 
\begin{align}
\mathcal{L}_{\rm NP}=~&-\frac{1}{4}C^{\mu\nu}C_{\mu\nu}+\bar{\chi}\left(i\slashed{\mathcal{D}}-m_{\chi}\right)\chi+(D^\mu S_1)^\dagger(D_\mu S_1)+(\mathcal{D}^\mu \phi)^\dagger(\mathcal{D}_\mu \phi)\nonumber\\
&-\Big[Y_L^{j2} (\bar{Q}_L^{Cja}\theta^{ab}L_L^{2b})S_1+ Y_R^{j2} 
(\bar{U}_R^{Cj}\mu_R)S_1+{\rm h.c.}\Big]~.
\label{eq:Lag1}
\end{align}
Here, $C_{\mu\nu}=\partial_\mu Z^\prime_\nu-\partial_\nu Z^\prime_\mu$ denotes the field strength tensor associated with $U(1)_{L_e-L_\mu}$ gauge boson $Z^\prime$. $Y_{L}$ and $Y_R$ are completely arbitrary $3\times 3$ Yukawa matrices in the flavor basis. The superscript $C$ defines the charge-conjugation and $\{a,b\}$ stands for the $SU(2)_L$ indices. $j$ is the generation index for quarks while for leptons, only the $2^{\rm nd}$ generation is allowed due to the particular $U(1)_{L_e-L_\mu}$ charge assignment of $S_1$. $\theta^{ab}=(i\sigma_2)^{ab}$, $\sigma_k$~($k=1,\,2,\,3$) being the Pauli matrices. The covariant derivatives can be defined as,
\begin{align}
\mathcal{D}_\mu~&=\partial_\mu-ig^\prime Q^\prime Z^\prime_\mu~,\nonumber\\
D_\mu~&=\partial_\mu-ig_3\frac{T^n}{2}G_\mu^n-ig_1 Y B_\mu-ig^\prime Q^\prime Z^\prime_\mu~,
\end{align}
where, $Q^\prime$ denotes the $U(1)_{L_e-L_\mu}$ charge of a particular field. $G_\mu^n$~($n=1,\cdots, 8$) and $B_\mu$ represent the gauge bosons associated with $SU(3)_C$ and $U(1)_Y$, respectively, with $T^n$ being the Gell-Mann matrices. $g_k$~($k=1,2,3$) and $g^\prime$ label the SM and $U(1)_{L_e-L_\mu}$ gauge couplings, respectively. 

Using the freedom to rotate equal-isospin fermion fields in the flavor basis, one can assume the charged lepton and down-type quark Yukawas to be diagonal so that the transformation from flavor to mass basis is given by $u_L\to \left(\mathbf{V}_{\rm CKM}^\dagger\right) u_L$ and $\nu_L\to \left(\mathbf{U}_{\rm PMNS}\right) \nu_L$. Here $\mathbf{V}_{\rm CKM}$ and $\mathbf{U}_{\rm PMNS}$ represent the CKM and PMNS matrices, respectively. Therefore, in the physical basis Eq.~\eqref{eq:Lag1} can be recast as,   
\begin{align}
\mathcal{L}_{\rm NP}=~& -\frac{1}{4}C^{\mu\nu}C_{\mu\nu}+\bar{\chi}\left(i\slashed{\mathcal{D}}-m_{\chi}\right)\chi+(D^\mu S_1)^\dagger(D_\mu S_1)+(\mathcal{D}^\mu \phi)^\dagger(\mathcal{D}_\mu \phi)\nonumber\\
&-\Big[\left\{\bar{u}_L^{Cj}\left(\mathbf{V}^*_{\rm CKM}Y_L\right)^{j2} \mu_L\right\}S_1
-\left\{\bar{d}_L^{Cj}\left(Y_L \mathbf{U}_{\rm PMNS}\right)^{j2} (\nu_\mu)_L\right\}S_1+ Y_R^{j2} 
(\bar{U}_R^{Cj}\mu_R)S_1+{\rm h.c.}\Big]\nonumber\\
=~& -\frac{1}{4}C^{\mu\nu}C_{\mu\nu}+\bar{\chi}\left(i\slashed{\mathcal{D}}-m_{\chi}\right)\chi+(D^\mu S_1)^\dagger(D_\mu S_1)+(\mathcal{D}^\mu \phi)^\dagger(\mathcal{D}_\mu \phi)\nonumber\\
&-\Big[\left(\bar{u}_L^{Cj}\xi_L^{j2} \mu_L\right)S_1-\left(\bar{d}_L^{Cj}\beta^{j2} (\nu_\mu)_L\right)S_1+ \xi_R^{j2} 
(\bar{U}_R^{Cj}\mu_R)S_1+{\rm h.c.}\Big]~,
\label{eq:NP_Lag}
\end{align}
where, $\xi_L^{j2}=\left(\mathbf{V}^*_{\rm CKM}Y_L\right)^{j2}$, $\beta^{j2}=\left(Y_L \mathbf{U}_{\rm PMNS}\right)^{j2}$, and $\xi_R^{j2}=Y_R^{j2}$. However, the neutrino sector doesn't play any role in the present analysis. The vector-like fermion $\chi$ being stabilized through an exact $Z_2$ symmetry, can be considered as a viable DM candidate with mass $m_\chi$. Finally, the scalar potential $\mathbb{V}$, with only renormalizable couplings, can be read as,
\begin{align}
\mathbb{V}(H,\,S_1,\,\phi)=&~\mu_H^2(H^\dagger H)+\tilde{M}_{S_1}^2(S_1^\dagger S_1)+\mu_\phi^2(\phi^\dagger\phi)+\lambda_1(H^\dagger H)(S_1^\dagger S_1)+\lambda_2(H^\dagger H)(\phi^\dagger \phi)\nonumber\\
&+\lambda_3(S_1^\dagger S_1)(\phi^\dagger \phi)+\lambda_4(S_1^\dagger S_1)^2+\lambda_5(\phi^\dagger \phi)^2+\lambda_H(H^\dagger H)^2~.
\label{eq:V_pot}
\end{align}
Note that, after SSB, the term $\lambda_2(H^\dagger H)(\phi^\dagger \phi)$ generates the mixing between the SM Higgs and $\phi$. However, the Higgs sector is strongly constrained through various collider searches~\cite{CMS:2018amk, ATLAS:2018sbw} resulting in an upper bound on the mixing angle~\cite{Robens:2022cun}. Therefore, one can assume $\lambda_2$ to be negligibly small so that the presence of $\phi$ doesn't affect the Higgs observables. In Eq.~\eqref{eq:V_pot}, $\tilde{M}_{S_1}$ is the bare mass term for $S_1$. The SM Higgs acquires vacuum expectation value~(VEV) at the electroweak scale $\Lambda_{\rm EW}$, while $U(1)_{L_e-L_\mu}$ is broken spontaneously at a higher energy scale $\Lambda_{e\mu}$ where $\phi$ acquires a VEV $v_\phi$. Thus, after SSB, the scalar sector can be redefined as,
\begin{align}
H = 
\frac{1}{\sqrt{2}}\begin{pmatrix}
0 \\
h+v_H
\end{pmatrix}, \quad\quad \phi=\frac{1}{\sqrt{2}}(\phi+v_\phi), \quad\quad S_1=S_1,
\end{align}
with $v_H=246$ GeV being the EW VEV. Note that, $S_1$ being {\it color-protected}, can't acquire a VEV. In the broken phase of $U(1)_{L_e-L_\mu}$, $Z^\prime$  becomes massive with a mass $M_{Z^\prime}=g^\prime Q_\phi v_\phi$. Further, following the vacuum stability conditions~\cite{Kannike:2012pe}, the physical mass terms for the scalars can be obtained as,
\begin{align}
M^2_h=&~2\lambda_Hv_H^2~,\nonumber\\
M^2_\phi=&~2\lambda_5v_\phi^2~,\nonumber\\
M^2_{S_1}=&~\tilde{M}_{S_1}^2+\frac{\lambda_1}{2}v_H^2+\frac{\lambda_3}{2}v_\phi^2~,
\label{eq:mass}
\end{align}
where $M_h=125$ GeV stands for the SM Higgs mass. The collider searches restrict the LQ mass $M_{S_1}\geq \mathcal{O}(1)$ TeV~\cite{ParticleDataGroup:2022pth}. Moreover, one has to assume $M_\phi>M_h/2$ to evade any constraint from the invisible $h$ decay~\cite{ATLAS:2022yvh}.  

Note that the kinetic mixing strength between two abelian gauge bosons is a free parameter in the theory and, in principle, not forbidden by any symmetry. However, as $U(1)_{L_e-L_\mu}$ can be embedded into a higher non-abelian gauge group $SU(2)_{L_e-L_\mu}$~\cite{He:1991qd, Bauer:2018onh}, the kinetic mixing term vanishes at the tree-level. Moreover, the finiteness of the loop-induced kinetic mixing is guaranteed by the fact that $SU(2)_{L_e-L_\mu}$ breaks to $U(1)_{L_e-L_\mu}$ without any mixing with the $U(1)_Y$. The effect of loop-induced kinetic mixing has been discussed in the Refs.~\cite{Bauer:2018onh,Ballett:2019xoj,Ardu:2022zom, Dasgupta:2023zrh}.

\section{Dark Matter Phenomenology}
\label{sec:DM}
\noindent
As stated in Sec.~\ref{sec:model}, the observed dark matter relic density can be explained by introducing an additional vector-like SM-singlet fermion $\chi$, charged under $U(1)_{L_e-L_\mu}$ and $Z_2$. After SSB, the imposed $Z_2$ symmetry protects it against all the kinematically accessible decay channels. Note that, $Z^\prime$ is the only portal for $\chi$ to interact with the SM. Thus, from Eq.~\eqref{eq:Lag1}, $\bar{\chi}\chi Z^\prime$ interaction can be obtained as,
\begin{align}
\mathcal{L}_{\rm int}^\chi=~&g^\prime Q_\chi\bar{\chi}\gamma^\mu\chi Z^\prime_\mu=g_\chi\bar{\chi}\gamma^\mu\chi Z^\prime_\mu~.
\end{align} 
Though for a particular value of $M_{Z^\prime}$, $g^\prime$ is constrained through various experiments, $Q_\chi$ is an entirely free parameter. Thus, one can vary the effective $\bar{\chi}\chi Z^\prime$ coupling, $g_\chi=g^\prime Q_\chi$ between $\mathcal{O}(10^{-2})$ to $\mathcal{O}(1)$, such that the current DM abundance can be explained through the freeze-out mechanism.  
\begin{figure}[!ht]
\centering
\subfloat[(a)]{\includegraphics[scale=0.6]{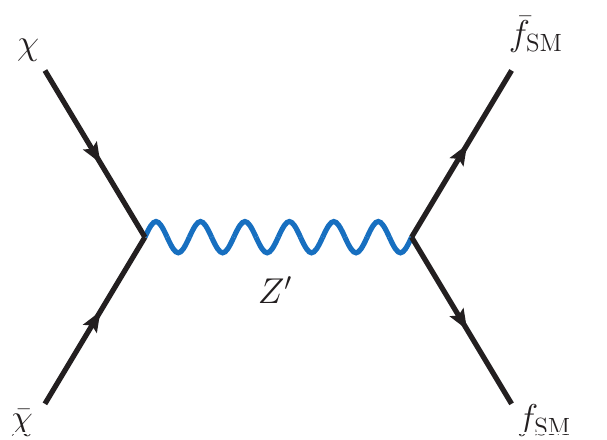}}\qquad\qquad\qquad\qquad
\subfloat[(b)]{\includegraphics[scale=0.6]{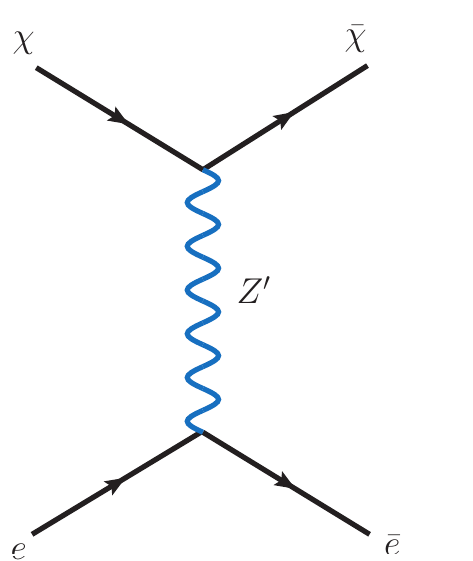}}
\caption{(a) Annihilation channels contributing to the DM relic density. For $m_\mu<m_\chi\leq 1$ GeV, $f_{\rm SM}=e,\mu,\nu_e,\nu_\mu$. (b) DM-electron scattering through the $Z^\prime$ boson.}
\label{fig:DM}
\end{figure}
Fig.~\ref{fig:DM}\,(a) represents all the possible $s$-channel diagrams contributing to the annihilation cross section of $\chi$. $Z^\prime$ being a leptophilic neutral gauge boson, $f_{\rm SM}$ can be $e,\mu,\nu_e,\nu_\mu$ as long as $m_\chi>m_\mu$. For $m_\chi\leq m_\mu$, $\bar{\chi}\chi\to\bar{\mu}\mu$ can't be accessed. Fig.~\ref{fig:DM}\,(b) depicts the DM-electron scattering process mediated by $Z^\prime$. $\chi$ being charged under $U(1)_{L_e-L_\mu}$ with $m_\chi\leq 1$ GeV, this particular scattering process is extremely significant to constraint the parameter space through the direct search experiments and DM annihilation at the white dwarfs. 
\subsection{Relic Density}
The vector-like fermion $\chi$, being a weakly interacting particle~[since, $g_\chi\sim\mathcal{O}(0.01-1)$], maintained a thermal equilibrium in
the early universe with the SM particles. However, as the universe evolved further, at a certain point in time, the interaction rate of $\chi$ with the SM particles fell short compared to the expansion rate of the universe, and $\chi$ decoupled from the thermal bath. Moreover, $\chi$ being protected against any decay to the SM fields, its abundance was frozen forever to the decoupling value, resulting in its current relic density, $\Omega_\chi h^2$. The value can be obtained by solving the Boltzmann equation:
\begin{align}
\frac{dn_\chi}{dt}+3\mathcal{H}n_\chi=-\left\langle\sigma_{\rm An}v\right\rangle\Big[n_\chi^2-(n_\chi^{\rm eq})^2\Big]~,
\label{eq:boltz}
\end{align}
where $n_\chi$ stands for the number density of $\chi$ with the superscript `eq' representing its equilibrium value. $\mathcal{H}$ defines the Hubble parameter, and $\left\langle\sigma_{\rm An}v\right\rangle$ is the thermal averaged annihilation cross section times the relative velocity~($v$) of $\chi$. Note that, since $m_\chi$ lies in the sub-GeV mass regime, to satisfy the observed relic density near the resonance funnel, one should assume $M_{Z^\prime}\sim\mathcal{O}(m_\chi)$. To be precise, $M_{Z^\prime}=200$ MeV will be considered for all the subsequent discussions. Eq.~\eqref{eq:boltz} has been solved numerically using $\mathtt{micrOMEGAs}$~\cite{Belanger:2010pz} with the model files generated through $\mathtt{LanHEP}$~\cite{Semenov:2008jy}. 
\begin{figure}[!ht]
\centering
\includegraphics[scale=1]{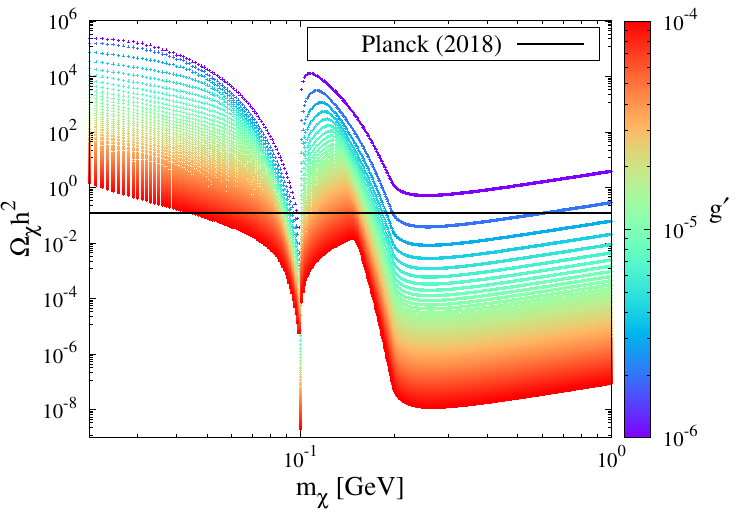}
\caption{The variation of relic density as a function of the DM mass $m_\chi$ for $M_{Z^\prime}=200$ MeV. The black solid line represents the observed DM abundance from Planck~\cite{Planck:2018vyg}. $Q_\chi=10^4$ has been considered for the analysis.}
\label{fig:relic}
\end{figure}
Fig.~\ref{fig:relic} shows the variation of $\Omega_\chi h^2$ as a function of $m_\chi$ with $g^\prime$ varying continuously between $10^{-6}$ to $10^{-4}$ in steps of $10^{-6}$. The black solid line marks the observed DM abundance, $\Omega_{\rm DM}h^2=0.1198\pm 0.0012$~\cite{Planck:2018vyg}. Therefore, only those parameter space points in the $m_\chi-g^\prime$ plane can represent a viable DM candidate for which $\Omega_\chi h^2=\Omega_{\rm DM}h^2$. Note that, the choice of $Q_\chi=g_\chi/g^\prime$ is entirely from a phenomenological perspective and physically implies that $Z^\prime$ shows a stronger interaction with $\chi$ compared to the SM leptons. Moreover, the assumed range of $g_\chi$ is consistent with the bounds from the ellipticity of the galactic DM halos which can be read as $g_\chi\leq 0.1\left(\frac{M_{Z^\prime}}{10~{\rm MeV}}\right)\left(\frac{m_\chi}{100~{\rm MeV}}\right)^{-1/4}$~\cite{Feng:2009hw,Essig:2011nj}. 
\subsection{Direct Detection through DM-Electron Scattering}
Direct detection experiments searching for DM-electron scattering are crucial to test the proposed model for two reasons --- $(i)$ $\chi$ being non-trivially charged under $U(1)_{L_e-L_\mu}$, can have tree-level interaction only with the first two lepton generations. In the proposed framework, DM-quark scattering will be naturally loop-suppressed, leading to a negligibly small nuclear scattering cross section. $(ii)$ The assumption that $\chi$ exists in the sub-GeV mass regime results in a maximal response when $\chi$ scatters on electrons. For a sub-GeV DM, the typical order of energy deposition to the target material varies from $1-10$ eV, which is well below the current sensitivity of the nuclear recoil experiments. However, it is large enough to trigger various atomic processes, e.g., electron ionization, electron excitation, and molecular dissociation. Currently, we have only feasible experimental facilities to probe the electron excitation signals that may arise from a DM scattering event on the regular matter. On an arbitrary target material, the DM scattering rate per unit detector mass can be formulated from the Fermi's Golden rule as~\cite{Trickle:2019nya, Kahn:2021ttr},
\begin{align}
R=\frac{\rho_\chi}{\rho_T\,m_\chi}\int d^3\vec{v}\,\,G_\oplus(\vec{v})\frac{V\,d^3\vec{p}_\chi^\prime}{(2\pi)^3}\sum_f\left|\left\langle f,\vec{p}_\chi^\prime\left|\mathbb{H}_{\chi T}\right|i,\vec{p}_\chi\right\rangle\right|^2 2\pi\delta(E_f-E_i+E^\prime_\chi-E_\chi)~,
\label{eq:rate1}
\end{align}
where $V$ and $\rho_T$ are the volume and density of the detector, respectively. $G_\oplus(\vec{v})$ defines the DM velocity distribution in the earth frame of reference with $\rho_\chi=0.4$ GeV/cm$^3$ being the local DM density~\cite{Catena:2009mf,Salucci:2010qr}. $p_\chi=(E_\chi,\,\vec{p}_\chi)$ and $p^\prime_\chi=(E_\chi^\prime,\,\vec{p}_\chi^\prime)$ define the 4-momenta of the incoming and outgoing DM particle $\chi$, whereas $|i\rangle$~($E_i$) and $|f\rangle$~($E_f$) denote the initial and final states~(energies) of the target, respectively. Note that, in general, the detector is assumed to be initially in the ground state at zero temperature. Thus, Eq.~\eqref{eq:rate1} doesn't contain a sum over the ensemble of initial states. The interaction between $\chi$ and the target material is described by the non-relativistic~(NR) Hamiltonian $\mathbb{H}_{\chi T}$ which can be treated as a perturbation term on the free DM Hamiltonian. The assumption ensures that the unperturbed eigenstates are described by plane wave solutions, and also forbids any entanglement between $\chi$ and the detector states, i.e., $|i,\vec{p}_\chi\rangle=|i\rangle\otimes|\vec{p}_\chi\rangle$ and $|f,\vec{p}_\chi^\prime\rangle=|f\rangle\otimes|\vec{p}_\chi^\prime\rangle$. 

In the present framework, DM-target interaction follows from a single effective operator. Therefore, the matrix elements in Eq.~\eqref{eq:rate1} can be decomposed as, 
\begin{align}
\left\langle f,\vec{p}_\chi^\prime\left|\mathbb{H}_{\chi T}\right|i,\vec{p}_\chi\right\rangle=\int \frac{d^3\vec{q}}{(2\pi)^3}\left\langle\vec{p}_\chi^\prime\left|\mathcal{A}_\chi(\vec{q})\right|\vec{p}_\chi\right\rangle\times\left\langle f\left|\mathcal{A}_T(\vec{q})\right|i\right\rangle~,
\label{eq:rate2}
\end{align} 
where $\mathcal{A}_T$ and $\mathcal{A}_\chi$ signify the operators acting only on the target system and DM states, respectively. $\vec{q}=\vec{p}_\chi^\prime-\vec{p}_\chi$ represents the momentum transfer due to scattering. Using plane wave approximation for the DM states, Eq.~\eqref{eq:rate2} results in~\cite{Kahn:2021ttr},
\begin{align}
\left\langle f,\vec{p}_\chi^\prime\left|\mathbb{H}_{\chi T}\right|i,\vec{p}_\chi\right\rangle=\frac{1}{V}\sqrt{\frac{\pi\bar{\sigma}(q)}{\mu_{\chi T}^2}}\,\,\left\langle f\left|\mathcal{A}_T(\vec{q})\right|i\right\rangle~.
\label{eq:matrix}
\end{align}
Here, $q=|\vec{q}|$ and $\mu_{\chi T}=(m_\chi m_T)/(m_\chi+ m_T)$ defines the reduced mass of the DM-target 2-body system, with $m_T$ being the mass of the target particle. For DM-electron scattering, $m_T=m_e$. The interaction strength, specific to a DM model, is encapsulated in the effective cross section $\bar{\sigma}(q)$. However, following the usual convention $\bar{\sigma}(q)$ can be formulated as,
\begin{align}
\bar{\sigma}(q)=\bar{\sigma}_e\left|\mathcal{F}_{\rm DM}(q)\right|^2~,
\end{align}
where $\bar{\sigma}_e$ defines the effective DM-electron elastic scattering cross section. From Fig.~\ref{fig:DM}\,(b), it can be calculated as~\cite{Essig:2011nj},
\begin{align}
\bar{\sigma}_e=\frac{\mu_{\chi e}^2(g^\prime)^4Q_\chi^2}{\pi(\alpha^2m_e^2+M_{Z^\prime}^2)^2}~,
\label{eq:sig_e}
\end{align}
with $\alpha$ denoting the fine structure constant. Note that, $\bar{\sigma}_e$ can be interpreted as the cross section for $\chi$ scattering off a free electron with reference momentum $q=\alpha m_e\simeq 3.7$ keV, after absorbing the momentum dependence of $\bar{\sigma}(q)$ into the DM form factor $\mathcal{F}_{\rm DM}(q)=(\alpha^2m_e^2+M_{Z^\prime}^2)/(q^2+M_{Z^\prime}^2)$. At this point, the DM-specific part of the interaction Hamiltonian can be completely described through $\bar{\sigma}(q)$. However, it's useful to further factorize Eq.~\eqref{eq:rate1} to separate out the target system part. Combining with Eq.~\eqref{eq:matrix} and introducing an auxiliary integration variable $\omega$, Eq.~\eqref{eq:rate1} can be recast as,
\begin{align}
R=&~\frac{\pi\rho_\chi\bar{\sigma}_e}{\rho_T\,m_\chi\,\mu_{\chi e}^2}\int d^3\vec{v}\,\,G_\oplus(\vec{v})\frac{d^3\vec{q}}{(2\pi)^3}\, d\omega\,\,\delta(\omega+E_\chi^\prime-E_\chi)\left|\mathcal{F}_{\rm DM}(q)\right|^2\nonumber\\
&\qquad\qquad\qquad\qquad\qquad\qquad\qquad\times\frac{2\pi}{V}\sum_f\left|\left\langle f\left|\mathcal{A}_T(\vec{q})\right|i\right\rangle\right|^2\delta(E_f-E_i-\omega)\nonumber\\
=&~\frac{\pi\rho_\chi\bar{\sigma}_e}{\rho_T\,m_\chi\,\mu_{\chi e}^2}\int d^3\vec{v}\,\,G_\oplus(\vec{v})\frac{d^3\vec{q}}{(2\pi)^3}\, d\omega\,\,\delta(\omega+E_\chi^\prime-E_\chi)\left|\mathcal{F}_{\rm DM}(q)\right|^2\times \mathbb{S}(\vec{q},\,\omega)~.
\label{eq:rate3}
\end{align}
The entire response of a particular target system towards the DM interaction is now contained within the {\it dynamic structure factor} $\mathbb{S}(\vec{q},\,\omega)$, where the normalization term $1/V$ indicates that $\mathbb{S}(\vec{q},\,\omega)$ is an intrinsic quantity. Since in the present work, $\chi$ interacts only with the electrons in an atomic or solid state detector system, the target-specific effective operator can be defined as, $\mathcal{A}_T(\vec{q})=-\sum_k e^{i\vec{q}.\vec{r}_k}$~\cite{Kaxiras_Joannopoulos_2019,Kahn:2021ttr}. Here, $\vec{r}_k$ denotes the position vector of the $k^{\rm th}$ electron. Thus, the dynamic structure factor reduces to
\begin{align}
\mathbb{S}(\vec{q},\,\omega)=\frac{2\pi}{V}\sum_f\left|\left\langle f\left|\sum_k e^{i\vec{q}.\vec{r}_k}\right|i\right\rangle\right|^2\delta(E_f-E_i-\omega)~.
\end{align} 
The detector states $|i\rangle$, $|f\rangle$ correspond to a generic many-body system and, in general, can't be momentum eigenstates. Since the analysis majorly focuses on spin-independent NR scattering events, the response of the detector material to the DM can be redefined in terms of the dielectric function $\epsilon(\vec{q},\,\omega)$, which encodes the response of the material to the EM interactions. Therefore, following Ref.~\cite{PhysRev.113.1254}, one can obtain,
\begin{align}
\mathbb{S}(\vec{q},\,\omega)&=\frac{q^2}{2\pi\alpha}\,{\rm Im}\left[-\frac{1}{\epsilon(\vec{q},\,\omega)}\right]~.
\label{eq:ELF1}
\end{align}
Note that, if the initial detector state corresponds to a finite temperature $T$, Eq.~\eqref{eq:ELF1} would have an additional multiplicative factor $[1-\exp(-\omega/k_BT)]^{-1}$~\cite{Girvin_Yang_2019, Knapen:2021run,Hochberg:2021pkt}, with $k_B$ representing the Boltzmann constant. However, for $T\to 0$, the factor tends to 1, resulting in Eq.~\eqref{eq:ELF1}. The factor ${\rm Im}\left[-1/\epsilon(\vec{q},\,\omega)\right]$ is commonly known as the {\it energy loss function}~(ELF) because it parametrizes the rate of losing energy~($\omega$) and momentum~($\vec{q}$) by a charged particle traversing through the material medium. ELF is an experimentally measured quantity and can be used to compute the DM scattering rate without using any explicit expression for the electron wavefunction within a particular target system. There are two significant advantages of this approach: $(i)$ The non-negligible impact of in-medium screening on the DM scattering rate is automatically included~\cite{Knapen:2021run}; $(ii)$ In the material science literature, ELF is an extremely well-explored parameter. Thus, standard and tested tools exist to extract it for any considered many-body environment. 

\begin{figure}[!ht]
\centering
\includegraphics[scale=1]{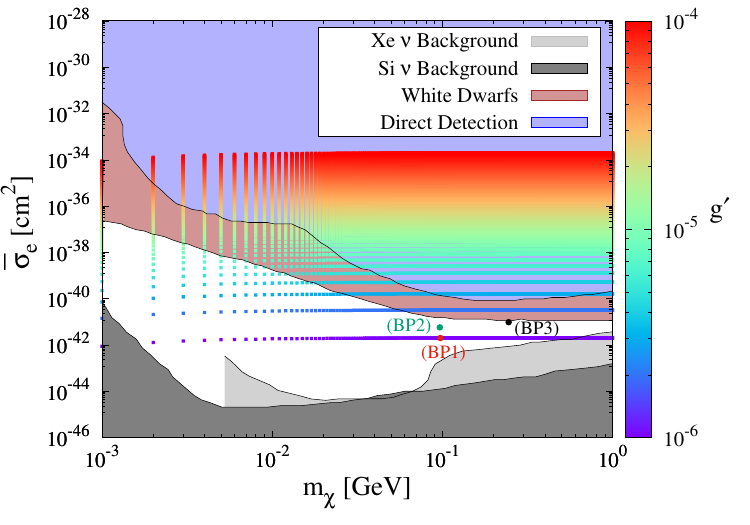}
\caption{The variation of DM-electron elastic scattering cross section as a function of $m_\chi$ for $M_{Z^\prime}=200$ MeV. As before, $Q_\chi=10^4$ has been considered for the analysis. The blue and brown-shaded portions represent the exclusion limits from direct detection experiments and DM annihilation at white dwarfs, respectively. The light and dark grey shades correspond to the irreducible neutrino background for Xe and Si-based detectors, respectively.}
\label{fig:sigma}
\end{figure}
Fig.~\ref{fig:sigma} displays the variation of $\bar{\sigma}_e$~[i.e., Eq.~\eqref{eq:sig_e}] as a function of the DM mass $m_\chi$ with $M_{Z^\prime}$ being fixed at 200 MeV. The continuous variation of $g^\prime$ from $10^{-6}$ to $10^{-4}$ has been depicted as before using the color palette. The light and dark grey shades in Fig.~\ref{fig:sigma} represent the {\it neutrino floor} corresponding to Xe and Si targets, respectively~\cite{Carew:2023qrj}. The blue shaded region is a compilation of the direct detection bounds from XENON1T~\cite{XENON:2019gfn, XENON:2019zpr}, DarkSide-50~\cite{DarkSide-50:2022qzh,DarkSide:2022knj}, PandaX-II~\cite{PandaX-II:2021nsg}, and SENSEI~\cite{SENSEI:2020dpa,SENSEI:2023zdf}. However, for sub-GeV DM, the most stringent bound on $\bar{\sigma}_e$ can be obtained from DM capture in the white dwarfs. The brown-shaded region defines the parameter space excluded through the observations of white dwarfs in the globular cluster M4~\cite{Dasgupta:2019juq, Bell:2021fye}.  

\begin{figure}[!ht]
\centering
\subfloat[\qquad(a)]{\includegraphics[scale=0.63]{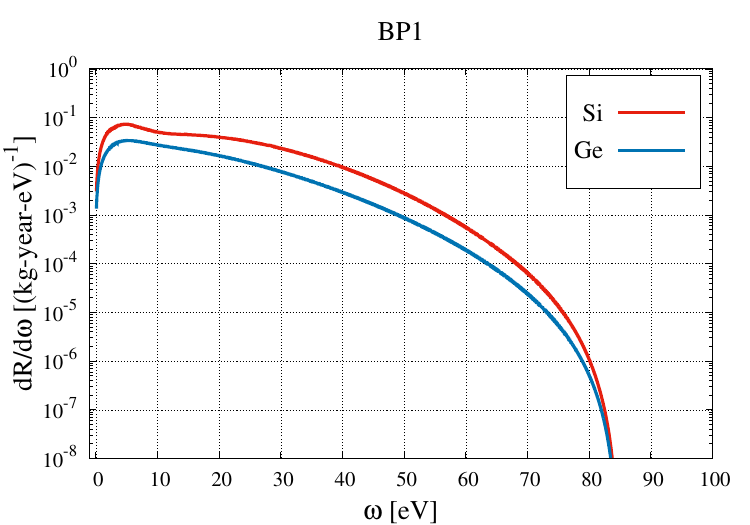}}\quad
\subfloat[\qquad(b)]{\includegraphics[scale=0.63]{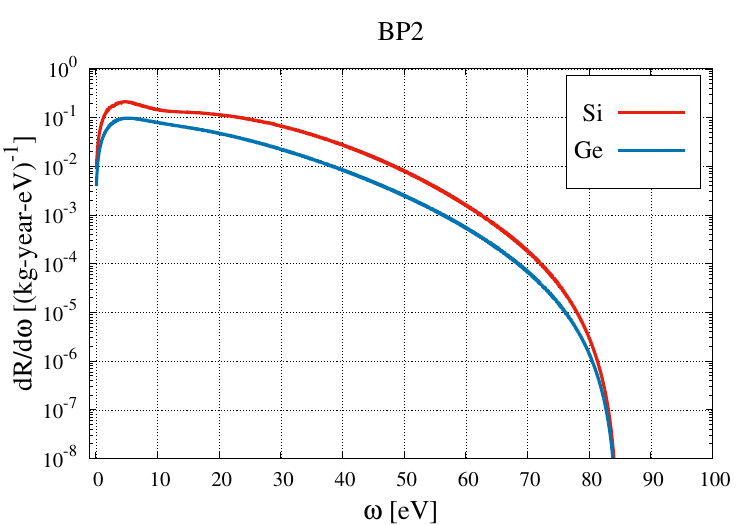}}\\
\subfloat[\qquad(c)]{\includegraphics[scale=0.63]{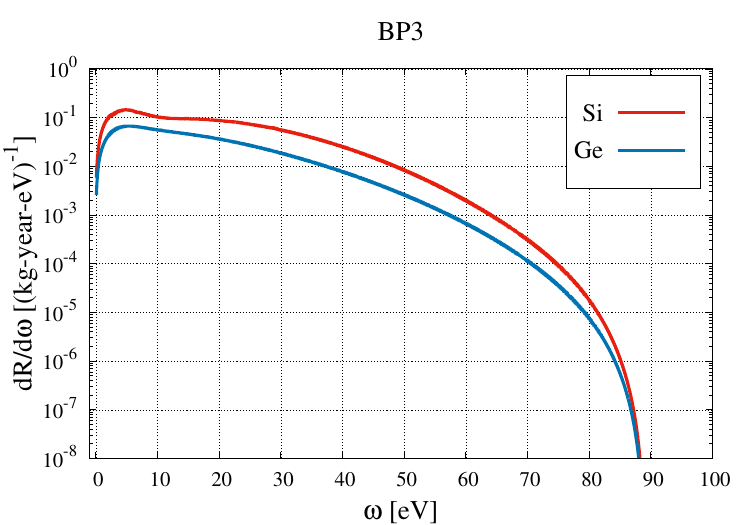}}
\caption{Variation of the differential DM-electron scattering rate $dR/d\omega$ as a function of the energy deposited~($\omega$) in the detector for (a) BP1, (b) BP2, and (c) BP3. The red and blue lines correspond to the Si and Ge-based detectors, respectively.}
\label{fig:rate}
\end{figure}

Table~\ref{tab:relic} enlists three benchmark points~(BP), which satisfy both relic density and direct detection constraints. 
\begin{table}[!ht]
\begin{tabular}{|c|c|c|c|}
\hline
Parameters & BP1 & BP2 & BP3\\
 \hline\hline
 $M_{Z^\prime}$ [MeV] & $200$ & $200$ & $200$ \\
 \hline
 $g^\prime$ & $1.00\times 10^{-6}$ & $1.30\times 10^{-6}$ & $1.48\times 10^{-6}$ \\
 \hline
 $m_\chi$ [MeV] & $97.37$ & $96.54$ & $245.00$ \\
 \hline
 $\Omega_\chi h^2$ & $0.119$ & $0.120$ & $0.121$\\
 \hline
 $\bar{\sigma}_e$ [${\rm cm}^2$] & $1.99\times 10^{-42}$ & $5.70\times 10^{-42}$ & $9.63\times 10^{-42}$\\
 \hline
\end{tabular}
\caption{Three benchmark points satisfying the observed DM relic density and direct detection bounds.}
\label{tab:relic}
\end{table}
In Fig.~\ref{fig:sigma}, these three benchmark points have been labeled with different colors: BP1~(red), BP2~(green), and BP3~(black); the same color code will be followed throughout the paper for representing them graphically. Fig.~\ref{fig:rate} depicts the variation of differential DM scattering rate for the benchmark points as a function of the deposited energy $\omega$. The analysis has been executed with a publicly available python-based numerical package, $\mathtt{DarkELF}$~\cite{Knapen:2021bwg}, which uses the energy loss function to determine DM-electron scattering rate on various possible target materials.
However, for the present purpose, only Si and Ge have been considered. Semiconductor targets are particularly useful for direct detection of sub-GeV DM due to their low bandgap energies.  The Mermin oscillator method has been used to generate the variations in Fig.~\ref{fig:rate}. It models the ELF as a weighted linear combination of the ELFs obtained with the Mermin dielectric function~\cite{Mermin:1970zz}. Moreover, the threshold has been set at $2e^-$ as for most of the experiments single electron bin may result in a large background. Approximately, the threshold to detect two ionization electrons corresponds to an energy deposition of 4.7 eV and 3.6 eV for Si and Ge, respectively~\cite{Knapen:2021run, Essig:2015cda, 10.1063/1.1656484}, where the differential event rate peaks. Note that, for a particular $\omega$, Fig.~\ref{fig:rate}\,(b), i.e., BP2 shows the highest scattering rate. For BP1, the suppression is due to a smaller value of $\bar{\sigma}_e$, whereas for BP3 it comes from a higher $m_\chi$. The event rate $R$ being inversely proportional to the DM mass, the enhancement due to a larger $\bar{\sigma}_e$ is nullified in the case of Fig.~\ref{fig:rate}\,(c). Moreover, the difference in densities for Si and Ge crystals is also reflected in the respective differential event rates. For cryogenic Si, $\rho_T=2.330$ ${\rm g/cm}^3$, whereas for Ge, it is 5.323 ${\rm g/cm}^3$~\cite{Knapen:2021bwg}. However, Ge target experiments can probe a comparatively lower mass range due to their smaller bandgap energy.

\section{Muon $\mathbf{(g-2)}$ : Reviving the $\mathbf{U(1)_{L_e-L_\mu}}$}
\label{sec:mu}
\noindent
The anomalous magnetic moment of muon~$\left(a_\mu\equiv (g-2)_\mu/2\right)$ has been determined with a very high degree of precision, both theoretically and experimentally. Thus, any discrepancy between the predicted and measured values of $a_\mu$ can be considered as a possible signature of some BSM physics. The current SM prediction for the anomalous magnetic moment of muon is given by $a_\mu^{\rm SM}= 116 591 810(43)\times 10^{-11}$~\cite{Aoyama:2020ynm}, whereas the 2023 experimental update from the Muon $g-2$ collaboration results in a world average of $a_\mu^{\rm Exp}=116 592 059(22)\times 10^{-11}$~\cite{Muong-2:2023cdq} at $5\sigma$ confidence level. Therefore, the observed discrepancy can be obtained as, 
\begin{align}
\Delta a^{\rm 2023}_\mu= a^{\rm Exp}_\mu- a^{\rm SM}_\mu=(2.49 \pm 0.48) \times 10^{-9}.
\label{eq:g2_2023}
\end{align} 
Note that, a recent lattice calculation of the hadronic vacuum polarization~(HVP) term by the BMW collaboration~\cite{Borsanyi:2020mff} and a preliminary experimental update from the CMD-3 detector~\cite{CMD-3:2023alj} indicate a notable tension with the present data which may result in a reduced and less significant discrepancy~\cite{Colangelo:2022jxc} between the predicted and observed values of $(g-2)_\mu$. However, at present, one can only use Eq.~\eqref{eq:g2_2023} to constrain the model parameters.
\subsection{Minimal $\mathbf{\mathcal{G}_{\rm SM}\otimes U(1)_{L_e-L_\mu}}$}
\begin{figure}[!ht]
\centering
\includegraphics[scale=0.6]{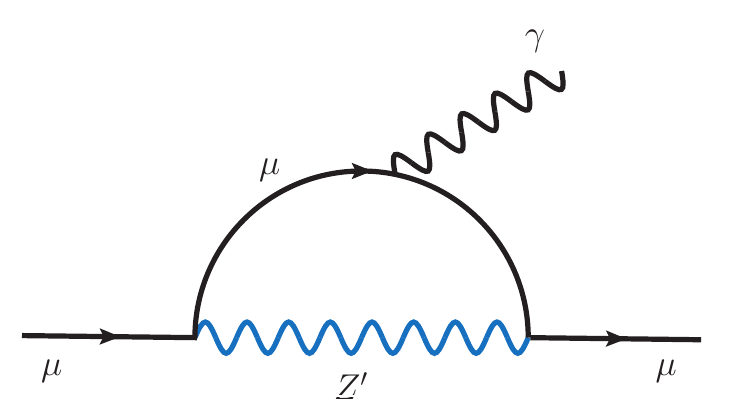}
\caption{One-loop correction to $\bar{\mu}\mu\gamma$ vertex within the minimal $\mathcal{G}_{\rm SM}\otimes U(1)_{L_e-L_\mu}$ theory.}
\label{fig:a1}
\end{figure}

\begin{figure}[!ht]
\centering
\includegraphics[scale=1]{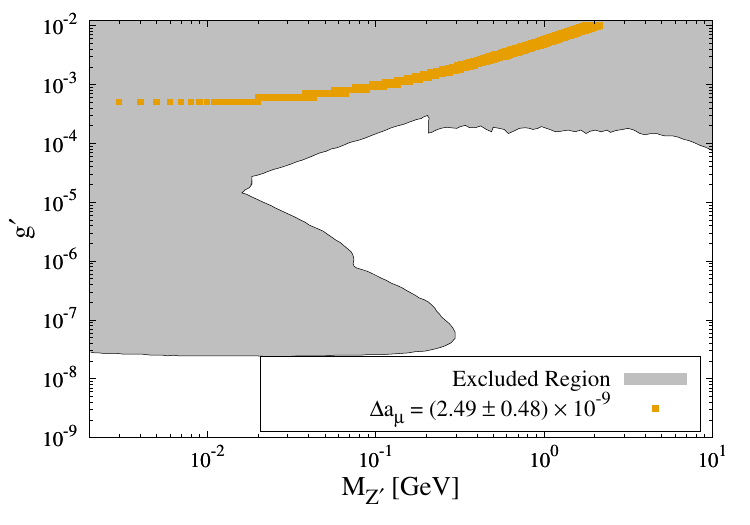}
\caption{The $M_{Z^\prime}-g^\prime$ plane for the minimal $\mathcal{G}_{\rm SM}\otimes U(1)_{L_e-L_\mu}$ model, with the golden dots signifying the parameter space points where $\Delta a^{\rm 2023}_\mu$ can be satisfied. The grey region denotes the experimentally excluded part of the parameter space. The exclusion limits have been compiled from Ref~\cite{Bauer:2018onh}.}
\label{fig:g2min}
\end{figure}

In the minimal $\mathcal{G}_{\rm SM}\otimes U(1)_{L_e-L_\mu}$ model, the only BSM contribution to $\bar{\mu}\mu\gamma$ vertex can be obtained at the one-loop level via $Z^\prime$, as shown in Fig.~\ref{fig:a1}. In the broken phase of $U(1)_{L_e-L_\mu}$, the corresponding NP correction to $(g-2)_\mu$ can be formulated as~\cite{Leveille:1977rc, Baek:2001kca},
\begin{align}
\Delta a_\mu=\Delta a_\mu^{(1)}=&~\frac{(g^\prime)^2}{8\pi^2}\int_0^1 dx\Bigg[\frac{2m_\mu^2 x^2(1-x)}{x^2m_\mu^2+(1-x)M_{Z^\prime}^2}\Bigg]\nonumber\\
=&~\frac{(g^\prime)^2}{8\pi^2}\int_0^1 dx\Bigg[\frac{2r x^2(1-x)}{x^2r+(1-x)}\Bigg]~,
\label{eq:a1}
\end{align}
where, $r=(m_\mu/M_{Z^\prime})^2$. 
Eq.~\eqref{eq:a1} has been solved numerically with $M_{Z^\prime}$ and $g^\prime$ varying continuously between 1 MeV$-$10 GeV and $10^{-8}-10^{-2}$, respectively. The golden patch in Fig.~\ref{fig:g2min} displays the parameter space where $\Delta a^{\rm 2023}_\mu$ can be explained with Eq.~\eqref{eq:a1}, whereas the grey-shaded region shows the parameter space that has already been probed through various experiments, e.g, electron beam dump experiments~\cite{Riordan:1987aw,Bjorken:1988as, Davier:1989wz, Bross:1989mp, Bjorken:2009mm, Andreas:2012mt}, electron fixed target experiments~\cite{APEX:2011dww,A1:2011yso, NA64:2016oww}, proton beam dump experiments~\cite{CHARM:1985nku,LSND:1997vqj,Blumlein:2011mv, Blumlein:2013cua}, collider searches~\cite{BaBar:2009lbr, BaBar:2014zli,KLOE-2:2011hhj,KLOE-2:2012lii,KLOE-2:2016ydq,Anastasi:2015qla}, neutrino experiments~\cite{Kaneta:2016uyt,Bilmis:2015lja,Lindner:2018kjo,Super-Kamiokande:2011dam,Ohlsson:2012kf,Gonzalez-Garcia:2013usa}, and white dwarf cooling~\cite{Dreiner:2013tja}. Moreover, the precise measurement of $N_{\rm eff}$~(relativistic degrees of freedom) at the time of the CMB formation is also vital to exclude the parameter space corresponding to $M_{Z^\prime}\leq \mathcal{O}(10)$ MeV~\cite{Ghosh:2024cxi}. Therefore, in the absence of any positive signal for a hidden gauge sector in the grey region of Fig.~\ref{fig:g2min}, the minimal $\mathcal{G}_{\rm SM}\otimes U(1)_{L_e-L_\mu}$ framework has been essentially ruled out as a viable explanation for the observed non-zero $\Delta a_\mu$.  
\subsection{Extended $\mathbf{\mathcal{G}_{\rm SM}\otimes U(1)_{L_e-L_\mu}}$}
\begin{figure}[!ht]
\centering
\subfloat[(a)]{\includegraphics[scale=0.6]{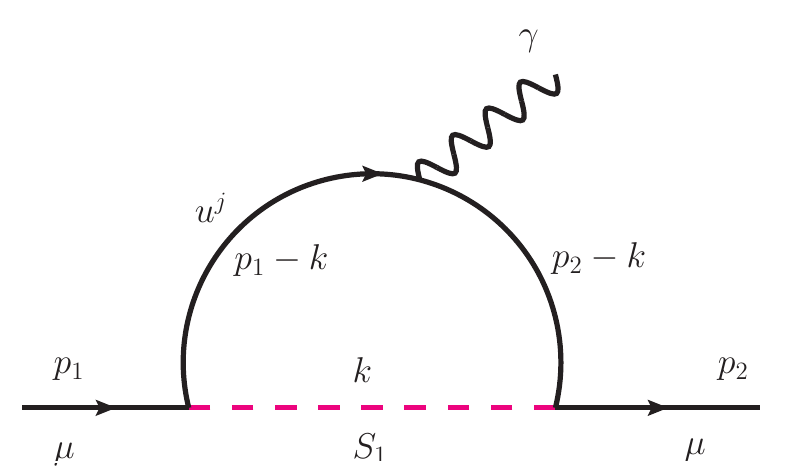}}\,\,
\subfloat[(b)]{\includegraphics[scale=0.6]{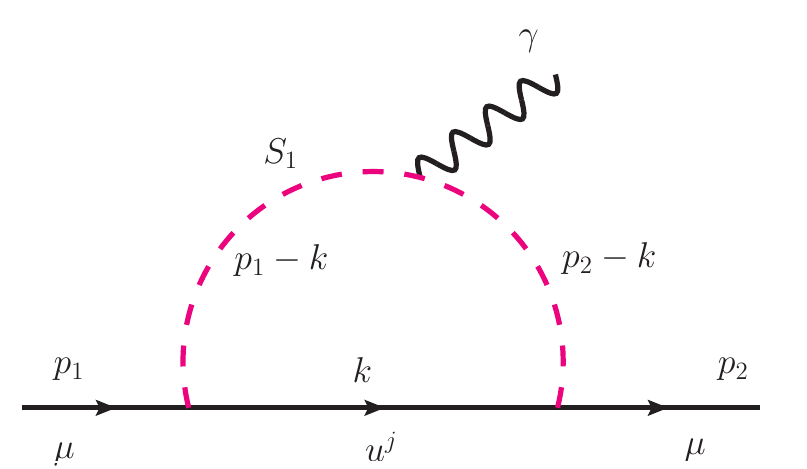}}
\caption{BSM corrections to the $\bar{\mu}\mu\gamma$ vertex in the extended $\mathcal{G}_{\rm SM}\otimes U(1)_{L_e-L_\mu}$ scenario, where (a) the up-type quarks couple to the photon~(Type-1 diagram), and (b) the LQ $S_1$ couples to the photon~(Type-2 diagram). $p_1,\,p_2$ define the external momenta.}
\label{fig:extg2}
\end{figure}

The proposed model extends the minimal $\mathcal{G}_{\rm SM}\otimes U(1)_{L_e-L_\mu}$ particle spectrum with a scalar LQ $S_1$. Due to the specific charge assignment~[see Table~\ref{tab:parti}], $S_1$ couples to muon and the up-type quarks, resulting in the one-loop corrections to the $\bar{\mu}\mu\gamma$ vertex as shown in Fig.~\ref{fig:extg2}. Fig.~\ref{fig:extg2}\,(a) shows the case where the photon couples to the up-type quarks, while Fig.~\ref{fig:extg2}\,(b) represents the situation when the photon touches the $S_1$ propagator~(magenta line). The former will be referred to as Type-1 diagram, while the latter will be called Type-2 for convenience.
\subsubsection*{Type-1 Diagram} The correction term to $\bar{\mu}\mu\gamma$ vertex due to the Type-1 diagram can be calculated as,
\begin{align}
\Delta \Gamma^\nu_1&= i\mathcal{N}_C\sum_{j\,=\,u,\,c,\,t}\int\frac{d^4k}{(2\pi)^4}\Bigg[\left(\xi_L^{j2}P_L+\xi_R^{j2}P_R\right)\frac{(\slashed{p}_2-\slashed{k}+m_j)}{(k-p_2)^2-m_j^2}(Q_{\rm EM}^j\gamma^\nu)\frac{(\slashed{p}_1-\slashed{k}+m_j)}{(k-p_1)^2-m_j^2}\nonumber\\
&\qquad\qquad\qquad\qquad\qquad\qquad\qquad\qquad\qquad\times\frac{1}{k^2-M_{S_1}^2}\left\lbrace(\xi_L^{j2})^*P_R+(\xi_R^{j2})^*P_L\right\rbrace\Bigg]\nonumber\\
&\equiv i\mathcal{N}_C\sum_{j\,=\,u,\,c,\,t}\int\frac{d^4k}{(2\pi)^4}\,Q_{\rm EM}^j\,\Bigg[\frac{\mathbb{N}_1^\nu}{\mathbb{D}_1}\Bigg].
\label{eq:t1}
\end{align}
Here, $\mathcal{N}_C=3$ denotes the color multiplicity factor, and $Q_{\rm EM}^j=2/3$ represents the EM charge for the up-type quarks in the unit of $e$, i.e., the charge of a proton. $m_j$ stands for the physical mass of the up-type quarks. The numerator of Eq.~\eqref{eq:t1} can be rearranged as,
\begin{align}
\mathbb{N}_1^\nu=\frac{1}{2}\Bigg[\mathcal{B}^j_1\Big\{&(\slashed{p}_2-\slashed{k})\gamma^\nu(\slashed{p}_1-\slashed{k})+m_j^2\gamma^\nu\Big\}+\mathcal{B}^j_2m_j\Big\{(\slashed{p}_2-\slashed{k})\gamma^\nu+\gamma^\nu(\slashed{p}_1-\slashed{k})\Big\}\nonumber\\
&+ \mathcal{B}^j_3\gamma^5\Big\{(\slashed{p}_2-\slashed{k})\gamma^\nu(\slashed{p}_1-\slashed{k})+m_j^2\gamma^\nu\Big\}+\mathcal{B}^j_4\,m_j\gamma^5\Big\{(\slashed{p}_2-\slashed{k})\gamma^\nu+\gamma^\nu(\slashed{p}_1-\slashed{k})\Big\}\Bigg],
\end{align}
where,
\begin{align}
\mathcal{B}^j_1&=\left|\xi_R^{j2}\right|^2+\left|\xi_L^{j2}\right|^2\,,\qquad\qquad\mathcal{B}^j_2=2\,{\rm Re} \left[(\xi_L^{j2})^*\xi_R^{j2}\right],\nonumber\\
\mathcal{B}^j_3&=\left|\xi_R^{j2}\right|^2-\left|\xi_L^{j2}\right|^2\,,\qquad\qquad\mathcal{B}^j_4=2i\,{\rm Im} \left[(\xi_L^{j2})^*\xi_R^{j2}\right].
\end{align}
After Feynman parametrization, the denominator can be recast as,
\begin{align}
\mathbb{D}_1=\Big[t^2-\Delta_1(x)\Big]^3,
\end{align}
where $t=k-yp_1-zp_2$ and $\Delta_1(x)=M_{S_1}^2\Big[x+\eta_j(1-x)\Big]$. $x,\,y,\,z$ are the Feynman parameters and $\eta_j=(m_j/M_{S_1})^2$. Note that, the calculation assumes an on-shell photon. Moreover, one can easily neglect the muon mass concerning the NP scale involved in the one-loop process. Integrating over the loop momentum $t$, the Type-1 contribution to the muon anomalous magnetic moment can be defined as,
\begin{align}
\Delta a_\mu^{(2a)}&=\frac{1}{8\pi^2}\sum_{j\,=\,u,\,c,\,t}\Bigg[\mathcal{B}^j_1\left(\frac{m_\mu}{M_{S_1}}\right)^2F_1(\eta_j)+\mathcal{B}^j_2\left(\frac{m_\mu\,m_j}{M^2_{S_1}}\right)F_2(\eta_j)\Bigg]~,
\label{eq:a2}
\end{align}
where, the functions $F_1$ and $F_2$ are given by,
\begin{align}
F_1(\kappa)&=\int_0^1\left[\frac{x(1-x)^2}{x+(1-x)\kappa}\right]\,dx=\frac{2+3\kappa-6\kappa^2+\kappa^3+6\kappa\,{\rm ln}\,\kappa}{6(1-\kappa)^4}~,\nonumber\\
F_2(\kappa)&=\int_0^1\left[\frac{(1-x)^2}{x+(1-x)\kappa}\right]\,dx=\frac{-3+4\kappa-\kappa^2-2\,{\rm ln}\,\kappa}{2(1-\kappa)^3}~.
\end{align}

\subsubsection*{Type-2 Diagram} $S_1$ being electromagnetically charged Fig.~\ref{fig:extg2}\,(b) arises as another contribution to the $\bar{\mu}\mu\gamma$ vertex. The corresponding correction term can be obtained as,
\begin{align}
\Delta \Gamma^\nu_2&= i\mathcal{N}_C\sum_{j\,=\,u,\,c,\,t}\int\frac{d^4k}{(2\pi)^4}\Bigg[\left(\xi_L^{j2}P_L+\xi_R^{j2}P_R\right)\frac{(\slashed{k}+m_j)}{k^2-m_j^2}.\frac{1}{(k-p_1)^2-M_{S_1}^2}\,.\,Q_{\rm EM}^{S_1}(p_1+p_2-2k)^\nu\nonumber\\
&\qquad\qquad\qquad\qquad\qquad\qquad\qquad\qquad\times\frac{1}{(k-p_2)^2-M_{S_1}^2}\left\lbrace(\xi_L^{j2})^*P_R+(\xi_R^{j2})^*P_L\right\rbrace\Bigg],\nonumber\\
&\equiv iQ_{\rm EM}^{S_1}\,\mathcal{N}_C\sum_{j\,=\,u,\,c,\,t}\int\frac{d^4k}{(2\pi)^4}\Bigg[\frac{\mathbb{N}_2^\nu}{\mathbb{D}_2}\Bigg]~.
\label{eq:t2}
\end{align}
Here $Q_{\rm EM}^{S_1}=1/3$ is the EM charge of $S_1$. Recasting the numerator of Eq.~\eqref{eq:t2}, one gets,
\begin{align}
\mathbb{N}_2^\nu=\frac{1}{2}\Big[\left\lbrace\mathcal{B}^j_1\slashed{k}+\mathcal{B}^j_2 m_j\right\rbrace+\gamma^5\left\lbrace\mathcal{B}^j_3\slashed{k}+\mathcal{B}^j_4 m_j\right\rbrace\Big] (p_1+p_2-2k)^\nu~.
\label{eq:lepg2_neu2}
\end{align}
Using Feynman parametrization, the denominator can be redefined as, 
\begin{align}
\mathbb{D}_2&=\Big[(k-yp_1-zp_2)^2-M_{S_1}^2\Big\{x\eta_j+(1-x)\Big\}\Big]^3\nonumber\\
    &=\Big[t^2-\Delta_2(x)\Big]^3.
\end{align}
Thus, the Type-2 diagram contributes to the muon anomalous magnetic moment as,
\begin{align}
\Delta a^{(2b)}_\mu &=-\frac{1}{16\pi^2}\sum_{j\,=\,u,\,c,\,t}\Bigg[\mathcal{B}^j_1\left(\frac{m_\mu}{M_{S_1}}\right)^2F_3(\eta_j)+\mathcal{B}^j_2\left(\frac{m_\mu\,m_j}{M^2_{S_1}}\right)F_4(\eta_j)\Bigg]~,
\label{eq:a3}
\end{align}
where,
\begin{align}
F_3(\kappa)&=\int_0^1\left[\frac{x(1-x)^2}{x\kappa+(1-x)}\right]\,dx=\frac{1-6\kappa+3\kappa^2+2\kappa^3-6\kappa^2\,{\rm ln}\,\kappa}{6(1-\kappa)^4}~,\nonumber\\
F_4(\kappa)&=\int_0^1\left[\frac{x(1-x)}{x\kappa+(1-x)}\right]\,dx=\frac{1 - \kappa^2 + 2\kappa\,{\rm ln}\,\kappa}{2 (1 - \kappa)^3}~.
\end{align}
Therefore, in the presence of a scalar LQ $S_1$, transforming non-trivially under $U(1)_{L_e-L_\mu}$, the leading order NP contribution to $(g-2)_\mu$ can be defined as,
\begin{align}
\Delta a_\mu=&~\Delta a_\mu^{(1)}+\Delta a_\mu^{(2)}\nonumber\\
=&~\Delta a_\mu^{(1)}+\Delta a^{(2a)}_\mu+\Delta a^{(2b)}_\mu~,
\label{eq:main}
\end{align}
where $\Delta a_\mu^{(1)}$ can be read from Eq.~\eqref{eq:a1}. Note that, for any set of parameter points, Fig.~\ref{fig:extg2}\,(a) and \ref{fig:extg2}\,(b) result in oppositely aligned contributions. Therefore, the sum $\left(\Delta a_\mu^{(2a)}+\Delta a_\mu^{(2b)}\right)$ must be so adjusted that one obtains $\Delta a_\mu=\Delta a_\mu^{\rm 2023}$ within allowed regions of $M_{Z^\prime}-g^\prime$ plane. 

In general, the NP couplings $\xi^{j2}_{L,\,R}$ can be complex and it is possible to predict  muon electric dipole moment~($\mu$EDM) and quark EDM within the considered framework. However, the analysis is beyond the scope of the paper. Therefore, for the subsequent discussions, $\xi^{j2}_{L,\,R}$ are assumed to be real so that $\mathcal{B}_4^j=0$.
\subsubsection*{Numerical Analysis}
In the extended $\mathcal{G}_{\rm SM}\otimes U(1)_{L_e-L_\mu}$ model, the parameter space relevant for describing the $(g-2)_\mu$, is given by,
\begin{align*}
\left\lbrace \xi^{j2}_L,\,\xi^{j2}_R,\, g^\prime,\, M_{Z^\prime},\, M_{S_1}\right\rbrace~,
\end{align*}
with $j=u,\,c,\,t$. Thus, there is a 9-dimensional parameter space in total. However, $M_{S_1}$ can be fixed at 2 TeV to simplify the numerical analysis without losing any physically significant outcome. The choice is completely in accordance with the collider bounds for $2^{\rm nd}$ generation scalar LQs~\cite{ParticleDataGroup:2022pth}. Moreover, in the absence of CLFV processes, the NP couplings $\xi^{j2}_L$ and $\xi^{j2}_R$ are unconstrained over the entire parameter space for all the quark generations. Therefore, to reduce the computational rigor, the flavor index can be suppressed, and one can assume $\xi^{j2}_L\equiv \xi_L$, and $\xi^{j2}_R\equiv \xi_R$. Note that, considering the quark-specific generation index will just enhance the degrees of freedom without any notable phenomenological effect. 
\begin{figure}[!ht]
\centering
\includegraphics[scale=1]{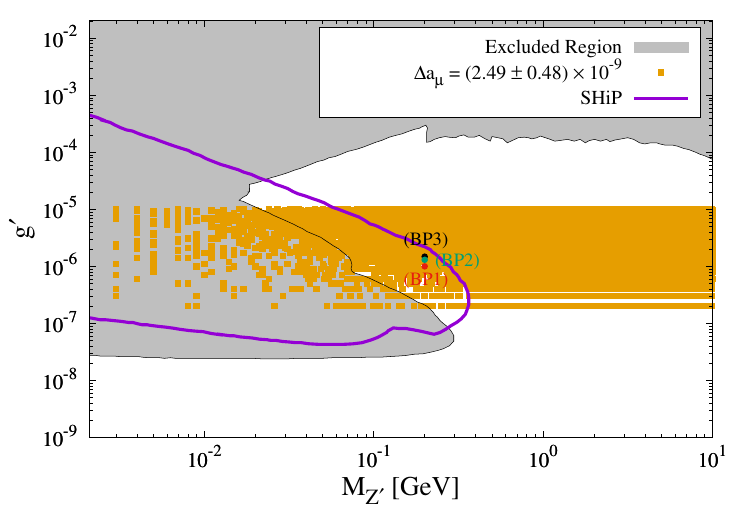}
\caption{The $M_{Z^\prime}-g^\prime$ plane corresponding to the extended $\mathcal{G}_{\rm SM}\otimes U(1)_{L_e-L_\mu}$ model. The golden dots represent the parameter space where $\Delta a^{\rm 2023}_\mu$ can be satisfied, whereas the grey-shaded region marks the experimentally excluded part of the parameter space. The violate line shows the future bound from the SHiP experiment. $M_{S_1}=2$ TeV has been assumed for the analysis.}
\label{fig:g2}
\end{figure}

Fig.~\ref{fig:g2} shows the $M_{Z^\prime}-g^\prime$ plane for the extended $\mathcal{G}_{\rm SM}\otimes U(1)_{L_e-L_\mu}$ framework. As before, the grey-shaded portion of the plot denotes the excluded regions of the parameter space. The crucial impact of having a scalar LQ charged under $U(1)_{L_e-L_\mu}$ is that the model can explain $\Delta a_\mu$ over the entire $M_{Z^\prime}-g^\prime$ plane. In case of a subdominant contribution from Fig.~\ref{fig:a1}, Fig.~\ref{fig:extg2} generates the compensating correction such that $\Delta a_\mu^{\rm 2023}$ can always be explained with Eq.~\eqref{eq:main}. However, for the present purpose, $g^\prime$ has been varied continuously from $10^{-7}$ to $10^{-5}$ for each $M_{Z^\prime}$ value between 1 MeV to 10 GeV. The LQ Yukawa couplings $\xi_L$ and $\xi_R$ have been varied randomly between $[-1,\,1]$ for each of the $(M_{Z^\prime},\,g^\prime)$ data point. The results have been shown with the golden dots in Fig.~\ref{fig:g2}. Following the aforementioned color convention, BP1, BP2, and BP3 have been marked with red, green, and black, respectively. Therefore, the three benchmark points can also explain the observed $\Delta a_\mu$ along with the DM phenomenology. Moreover, they are within the reach of the proposed SHiP experiment~\cite{SHiP:2015vad,Alekhin:2015byh,SHiP:2021nfo}. The projected bound has been outlined with violet in Fig.~\ref{fig:g2}.
\begin{figure}[!ht]
\centering
\includegraphics[scale=1]{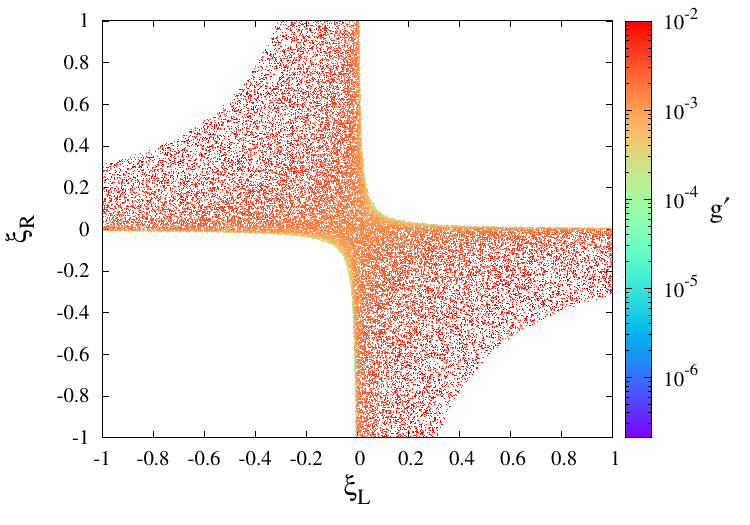}
\caption{$\Delta a^{\rm 2023}_\mu-$satisfying parameter points in the $\xi_L-\xi_R$ plane for different $g^\prime$ values with $M_{Z^\prime}=200$ MeV and $M_{S_1}=2$ TeV.}
\label{fig:scat}
\end{figure}

Fig.~\ref{fig:scat} displays a scatter plot in the $\xi_L-\xi_R$ plane. Here, $M_{Z^\prime}$ has been fixed at 200 MeV and $g^\prime$ varies continuously from $10^{-8}$ to $10^{-2}$. Note that, the range of $g^\prime$ has been chosen for a general purpose, and hence, the experimental bounds~[see Fig.~\ref{fig:g2}] have not been considered here. However, one can always extract the allowed parameter space from Fig.~\ref{fig:scat}. It's interesting to note that for smaller $g^\prime$ values~(approximately, $g^\prime\leq 10^{-4}$), ${\rm Re}\left[(\xi_L)^*\xi_R\right]\equiv \xi_L\xi_R>0$, whereas the product turns to be negative in the higher range of $g^\prime$. For $g^\prime\leq 10^{-4}$ with $M_{Z^\prime}=200$ MeV, Eq.~\eqref{eq:a1} results in $\Delta a_\mu^{(1)}\leq 1.24\times 10^{-11}$. Therefore, a major compensation must come from $\left(\Delta a_\mu^{(2a)}+\Delta a_\mu^{(2b)}\right)$ to explain the observed discrepancy. However, not all the terms of $\Delta a_\mu^{(2a)}$ and $\Delta a_\mu^{(2b)}$ contribute equally. It can be easily checked that for $0<\kappa<1$, the $F_i~[i=1,\,2,\,3,\,4]$ functions follow the hierarchy: $F_2(\kappa)>F_4(\kappa)>F_1(\kappa)>F_3(\kappa)$. Moreover, the first terms of Eqs.~\eqref{eq:a2} and \eqref{eq:a3} are suppressed by a factor of $(m_\mu/M_{S_1})^2\sim\mathcal{O}(10^{-9})$. Thus, the leading order contribution from $\left(\Delta a_\mu^{(2a)}+\Delta a_\mu^{(2b)}\right)$ effectively depends on the second term of Eq.~\eqref{eq:a2}, i.e., $\frac{1}{8\pi^2}\sum_{j\,=\,u,\,c,\,t}\Big[\mathcal{B}^j_2\left(\frac{m_\mu\,m_j}{M^2_{S_1}}\right)F_2(\eta_j)\Big]$. It clearly implies that for $\Delta a_\mu^{(1)}<\Delta a_\mu^{\rm 2023}$, $\mathcal{B}^j_2\equiv \mathcal{B}_2$ has to be positive, while for $\Delta a_\mu^{(1)}=\Delta a_\mu^{\rm 2023}$, $\mathcal{B}_2=0$. In practice, for the latter case, $\mathcal{B}_2=0$ corresponds to $\xi_L=\xi_R=0$ only. In case of $\Delta a_\mu^{(1)}>\Delta a_\mu^{\rm 2023}$, $\Delta a_\mu$ can only be set within the observed range if $\mathcal{B}_2<0$. Note that, with $\mathcal{B}_2=0$, it is impossible to create a negative contribution from $\left(\Delta a_\mu^{(2a)}+\Delta a_\mu^{(2b)}\right)$ due to the order of dominance between $F_1$ and $F_3$. However, only the first case, i.e., $\Delta a_\mu^{(1)}<\Delta a_\mu^{\rm 2023}$ is phenomenologically relevant; $\Delta a_\mu^{(1)}=\Delta a_\mu^{\rm 2023}$ and $\Delta a_\mu^{(1)}>\Delta a_\mu^{\rm 2023}$ correspond to the experimentally excluded regions of the $M_{Z^\prime}-g^\prime$ parameter space.
\section{Conclusions}
\label{sec:conc}
\noindent
The paper primarily addresses one of the major issues with all the $U(1)_{L_i-L_j}$ extensions of the SM. Though these well-motivated BSM theories provide a quite natural explanation for the observed discrepancies in lepton $(g-2)$, the associated parameter spaces have either been excluded or are on the verge of being probed through various experiments. The situation is particularly severe for $U(1)_{L_e-L_\mu}$. Thus, the proposed model extends the minimal particle spectrum of $\mathcal{G}_{\rm SM}\otimes U(1)_{L_e-L_\mu}$ theory with a scalar LQ $S_1$, transforming as $(\bar{\mathbf{3}},\,\mathbf{1},\,1/3,\,1)$ under the considered gauge group. The NP interaction between quarks and leptons is assumed to arise at the TeV scale. Such a theory has strong GUT motivation and is within the reach of the future high-energy colliders. Note that, the charge assignment of $S_1$ is specifically chosen to explain $\Delta a_\mu$ within the experimentally allowed parameter space. $\Delta a_e$ can also be explained using the same framework if one assigns a $U(1)_{L_e-L_\mu}$-charge of $-1$ to the LQ. However, for a simultaneous explanation of $\Delta a_\mu$ and $\Delta a_e$, the assumed particle spectrum has to be extended further as a single LQ charged under $U(1)_{L_e-L_\mu}$ can't couple to both electron and muon at the same time. The possibility is beyond the scope of the present work and can be explored in the future. In the considered model, $S_1$ can couple only the muon to the up-type quarks, leading to additional BSM contributions to $(g-2)_\mu$. These NP corrections compensate for the minimal $\mathcal{G}_{\rm SM}\otimes U(1)_{L_e-L_\mu}$ model as it produces a much-suppressed contribution to $(g-2)_\mu$ in the allowed parameter space. Thus, in the presence of a TeV-scale scalar LQ $S_1$, $\mathcal{G}_{\rm SM}\otimes U(1)_{L_e-L_\mu}$ gauge formulation can effectively be revived over the entire allowed region of $M_{Z^\prime}-g^\prime$ plane as a natural theory for $(g-2)_\mu$. Note that, though $U(1)_{L_e-L_\mu}$ has been considered here for analysis, the conclusions are generic and applicable for any of the $\mathcal{G}_{\rm SM}\otimes U(1)_{L_i-L_j}$ theory with a proper charge assignment of $S_1$. 

In addition to $\Delta a_\mu$, the model can also explain the observed DM abundance and the direct detection results. A $Z_2$-odd vector-like SM-singlet fermion $\chi$ with a non-zero $U(1)_{L_e-L_\mu}$ charge $Q_\chi$ has been introduced as a viable DM candidate. The model being naturally favorable for DM-electron scattering, $m_\chi$ has been restricted to the sub-GeV mass regime so that the DM-specific parameter space can be efficiently constrained through the direct detection and white dwarf bounds on $\bar{\sigma}_e$. Following the numerical results, three benchmark points have been chosen that satisfy both the relic density and direct detection constraints and thus represent testable parameter space points for the DM phenomenology. The actual detection prospects of $\chi$ have been predicted for Si and Ge-based detectors as they are more suitable for sub-GeV DM compared to the noble gas detectors. Due to band formation, semiconductors feature a better detection probability for light DM candidates. Out of the three benchmark points, BP2 is found to have a higher differential scattering rate. However, all three benchmark points satisfy $\Delta a_\mu^{\rm 2023}$ and correspond to a parameter space that can be probed with the future SHiP facility. Therefore, the direct detection experiments and searches for a hidden gauge sector can be used in complementarity of each other to test/falsify the model.

\bigskip
\small \bibliography{DM_elec}{}
\bibliographystyle{JHEPCust}    
    
\end{document}